\documentclass[12pt]{article} 
\title{Massless Particle Fields, with Momentum Matrices}  
\author{{\it Richard Shurtleff~}\thanks{affiliation and mailing 
address: Department of Applied Mathematics and Sciences, 
Wentworth Institute of Technology, 550 Huntington Avenue, 
Boston, MA, USA, ZIP 02115, telephone number: (617) 989-4338, fax 
number: (617) 989-4591 , e-mail address: shurtleffr@wit.edu}} 
\begin{document} 
          
\maketitle 

\begin{abstract}

Nontrivial translation matrices occur for spin $(A,B) \oplus (C,D)$ with $\mid A-C \mid$ = $\mid B-D \mid$ = 1/2, necessarily associating a $(C,D)$ field with a spin $(A,B)$ field. Including translation matrices in covariant non-unitary Poincar\'{e} representations also introduces new gauge terms in the construction of massless particle fields from canonical unitary fields. In the usual procedure without spacetime translation matrices, gauge terms arise from `translations' of the massless little group; the little group combines spacetime rotations and boosts making a group isomorphic with the Euclidean group $E_2,$ including $E_2$ translations. The usual remedy is to invoke gauge invariance. But here, the spacetime translation gauge terms can cancel the little group gauge terms, trading the need for gauge invariance with the need to specify displacements and to freeze two little group degrees of freedom that are not wanted anyway. The cancelation process restricts the helicity to $A-B-1$ for $A-C$ = $-(B-D)$ = 1/2 and $A-B+1$ for $A-C$ = $-(B-D)$ = $-1/2.$
 However, the cancelation only works for the little group standard momentum and specific transformations and, in general, gauge invariance is still needed to obtain massless particle fields. Expressions for massless particle fields for each spin type are found.
\end{abstract}

\section{Introduction} \label{Intro}

Successive infinitesimal rotations, boosts, and translations transform spacetime yet preserve the spacetime metric. The transformations can be represented canonically or covar-iantly.$^{ \cite{Wigner1}-\cite{W1}}$ Finite dimensional matrices in a canonical representations are unitary while finite dimensional matrices in a covariant representation are non-unitary. A covariant representation splits neatly into a matrix part which acts on `covariant' vectors to rearrange their components and a continuous part that acts to change the coordinates of the covariant vectors. The canonical representation cannot be split, the matrix part depends on the continuous part; the matrix part alone cannot represent spacetime transformations.

With canonical unitary representations, for momentum represented by $\bar{P},$ eigenvectors $a(p)$ can be found with eigenvalues $p,$ $\bar{P}a(p)$ = $pa(p).$ Hence, a translation shifts the phase of an eigenvector, $\exp{(-ix_{\mu}\bar{P}^{\mu})} a(p)$ = $\exp{(-ix_{\mu}p^{\mu})} a(p),$ where $\mu \in$ $\{1,2,3,4\}$ = $\{x,y,z,t\}.$ The massless class of vectors has a basis of eigenvectors with $p^2$ = 0; the massive class has a basis with $p^2 < 0.$ A canonical representation of any class can be founded upon a representation of a little group of transformations $\bar{W}$ that preserve a given standard momentum. For the massless class the standard momentum has null spacetime magnitude, say $k^{\mu}$ = $\{0,0,k^t,k^t\}.$ It can be shown that the massless little group is isomorphic to the group $E_2$ of rotations, $R(\theta),$ and translations, $S(\alpha,\beta),$ in an abstract two dimensional Euclidean plane. $^{ \cite{Wigner2}-\cite{W2}}$ 

With a covariant non-unitary representation, spacetime translations can change vector components something like rotations and boosts. But the character of a translation, whether a spacetime translation or a little group `translation' $S,$  differs from that of a rotation or boost.$^{ \cite{Mirman}}$ A matrix $D(1,\epsilon)$ representing a translation along spacetime displacement $\epsilon$ acts like a connection, $D(1,\epsilon) v$ = $v + \delta,$ i.e. inhomogeneous in $v.$  For example, consider the usual spin $(1/2,1/2)$ representation with spacetime 4-vectors. let $\Lambda^{\mu}_{\nu}$ be a homogeneous Lorentz transformation  and $b^{\mu}$ be a 4-vector displacement. Then the action of $\Lambda^{\mu}_{\nu}$ combined with translation by $b^{\mu}$ on coordinate vector $x^{\mu}$ can be represented$^{ \cite{Tung3},\cite{H}}$  as 
\begin{equation} \label{Lb1}
\pmatrix{\Lambda^{\mu}_{\nu} & b^{\mu} \cr 0 & 1} \pmatrix{x^{\nu} \cr 1} = \pmatrix{\Lambda^{\mu}_{\nu}x^{\nu} + b^{\mu} \cr 1} \quad .
\end{equation}
The diagonal block location of the rotations and boosts in this representation makes it clear that rotations and boosts, $\Lambda,$ give homogeneous results $\Lambda x.$ By putting $b$ in its place in the upper right triangular block of the transformation matrix, repeated application of the transformation cannot produce a term quadratic in $b.$ 
\begin{equation} \label{Lb2}
\pmatrix{\Lambda_1 & b_1 \cr 0 & 1} \pmatrix{\Lambda_2 & b_2 \cr 0 & 1}\pmatrix{x \cr 1} = \pmatrix{\Lambda_1 \Lambda_2 x +\Lambda_1 b_2 + b_1 \cr  1}  \quad .
\end{equation}
Thus, as in any similar representation, the translation by $b$ produces an inhomogeneous connection-like term.

With a covariant non-unitary representation, it is the connection-like action of the translation -like $S$ transformations of the little group that present obstacles to obtaining massless fields.$^{ \cite{W3}}$ But real spacetime translations are also part of the Poincar\'{e} group. The little group `translations' $S$ are actually specific combinations of rotations and boosts and are only isomorphic to $E_2$ translations. Thus little group translations occur in the diagonal blocks, see $\Lambda$ in (\ref{Lb1}) above, while real spacetime translations are generated by momenta in an off-diagonal triangular block; note  the location of $b$ in the above example.

In this paper, for covariant non-unitary representations,  the little group gauge terms from the translation-like $S$ are joined by gauge terms from spacetime translations. Two sets of nontrivial translation matrices,$^{ \cite{S}}$ denoted 12-  and 21-representations, exist for certain spin types, spin $(A,B)\oplus (C,D)$ with $\mid A-C \mid$ = $\mid B-D \mid$ = $1/2.$ The 12-representation is used herein. (Note that the above example, (\ref{Lb1}), has a 12-representation with spin $(1/2,1/2) \oplus (0,0)$ and the displacement $b$ located in the `12 block' of the matrix.) Given spacetime displacements $\epsilon^{\mu}$ and little group rotation angle $\theta$ one can determine the little group displacements $\alpha(\epsilon,\theta), \beta(\epsilon,\theta)$ that make the gauge terms vanish. The results can be applied to massless particle fields. 

A covariant vector $\psi$ can sometimes be constructed as a linear combination of canonical vectors $a,$ symbolically,
\begin{equation} \label{psi0}
\psi = \sum  u a  \quad ,
\end{equation}
where coefficients $u$ are called intertwiners. The vectors $a$ transform according to a canonical unitary representation; they are `single particle states.'  Let $D(\Lambda,b)$ indicate the covariant non-unitary representation of a spacetime transformation with homogeneous Lorentz transformation $\Lambda$ followed by a translation by a displacement $b.$ The covariant vector transforms as $\psi^{\prime}$ = $D \psi.$ Let $A$ indicate the corresponding canonical unitary representation of the same transformation, so that $a^{\prime}$ = $Aa.$ Then the transformation of the intertwiners is determined,
\begin{equation} \label{psi1}
\psi^{\prime} = D\psi =  \sum   Du a =  \sum  DuA^{-1} Aa = \sum   DuA^{-1} a^{\prime} \quad ,
\end{equation}
which implies that
\begin{equation} \label{u0}
u^{\prime} =  DuA^{-1}  \quad .
\end{equation}
When $u$ changes as in (\ref{u0}), there is no new information; the transformed relation (\ref{psi1}) simply repeats that the original untransformed linear relationship (\ref{psi0}) exists.  

The covariant vector $\psi$ is called a `particle field' when the intertwiners are invariant under spacetime transformations, $u^{\prime}$ = $u.$ Often, this assumption is made implicitly, as a part of the meaning of `particle field.'$^{ \cite{Tung4}-\cite{W4}}$ Thus the goal is to find invariant intertwiners $u$,  
\begin{equation} \label{u0a}
u^{\prime} =  DuA^{-1} = u  \quad ,
\end{equation}
so that the linear relationship is preserved with spacetime transformations,
\begin{equation} \label{psiPRIME1}
\psi^{\prime} = \sum  u a^{\prime} \quad . \end{equation} 
Whether or not there are invariant intertwiners depends in part on the class of the canonical representation.

For the massless class, there is always one gauge free field, exemplified by the standard vector $k^{\mu},$ with null spacetime magnitude $k_{\mu}k^{\mu}$ = 0. One can choose $u^{\prime}$ = $u$ = $e_0,$ where $e_0$ is a $k^{\mu}$-like vector. This field has helicity $A-B.$ There is no need to invoke gauge invariance or spacetime translations for this type of massless particle field. 

It is also shown in this paper  that, for `polarization-like' vectors $e_{\pm}$ a transformed intertwiner $u^{\prime},$ for momentum proportional to the standard momentum, can be a spacetime translated intertwiner, $D(1,\epsilon) u.$ This is not the desired invariance, $u^{\prime}$ = $u,$ unless $D(1,\epsilon)u$ is considered equivalent to $u,$ $D(1,\epsilon)u \cong $ $u$ implying that $u^{\prime} \cong $ $u.$ There is some heuristic justification for considering intertwiners at different points to be equivalent. Intertwiners are plane waves, $u \propto$ $\exp{(ipx)}$ and plane waves are invariant under translations $D(1,\epsilon)$ when the phase is invariant, $p \cdot \epsilon$ = 0. Thus the coordinate dependence of covariant representations has an equivalence relation for the intertwiners, and this offers some justification for extending the idea to the matrix part of covariant non-unitary representations. 

However, for general momenta, the equivalence involves gauge invariance. Thus even though gauge terms can be canceled for some momenta, the process does not generalize to gauge cancelation for all momenta.

The use of a displacement equivalence for fields and the gauge term cancelation restricts the helicity of the $(A,B)$ part of the field to $A-B \pm 1,$ which gives $\pm 1$ for photons, which have spin $(A,B)$ = $(1/2,1/2).$ Gravitons, spin (1,1), also would have the same helicity, $\pm 1,$ and not the extreme values expected, that is, $\pm 2.$ Whether or not forcing gravitons to have helicity $\pm 1$ disagrees with observation may be discussed elsewhere. Constraining the little group degrees of freedom, $\alpha, \beta,$ agrees with the observation that such degrees of freedom are not observed for photons.$^{ \cite{VDBij}- \cite{W5}}$  

For convenience and to set the notation, Section (\ref{P}) and Appendix (\ref{SMM}) describe the covariant non-unitary finite dimensional matrix representations of the Poincar\'{e} algebra used in this paper. In Sec.~(\ref{LG}), gauge terms are shown to appear with translation-like little group transformations applied to certain vectors. Similar gauge terms appear with spacetime translations along specified displacements in Sec.~(\ref{ST}). In Section (\ref{LGST}), by constraining the little group `displacements,' the gauge terms can cancel for `polarization-like' vectors $e_{\pm}.$ The application to the construction of massless particle fields in Sec.~(\ref{CCF}) finds massless fields for the four allowed spin types; just one helicity, $A-B \pm 1$ or $A-B ,$ occurs for each type. In Sec.~(\ref{PW}) equivalences are discussed that attempt to justify the change in the intertwiners upon transformation by relating the changes to the equivalences seen for plane waves. A problem set is included in Appendix (\ref{Pb}).

\section{Poincar\'{e} Matrix Generators } \label{P}

Nonzero momentum matrices occur for reducible angular momentum representations.$^{ \cite{S}}$ General results can be inferred from the spin $(A,B) \oplus (C,D)$ representation. By a suitable similarity transformation, if needed, the angular momentum and boost matrices of spin $(A,B) \oplus (C,D)$ can be put in block matrix form with the $AB$ block components given by the expressions in Appendix and the $CD$ block components given by the same expressions with $AB \rightarrow$ $CD,$
\begin{equation} \label{JABJCD} D(J_{i}) = \pmatrix{(J^{ (A,B)}_{i 11})_{ab,a_1 b_1} && 0 \cr 0 && (J^{ (C,D)}_{i 22})_{cd,c_1 d_1}} 
 \end{equation}
$$   D(K_{i}) = \pmatrix{(K^{ (A,B)}_{i 11})_{ab,a_1 b_1} && 0 \cr 0 && (K^{ (C,D)}_{i 22})_{cd,c_1 d_1}} \quad ,$$ where $i \in$ $\{1,2,3\}$ $= \{x,y,z\}$ and $D$ indicates a matrix representation. The notation for the matrix representing a generator, for example $D(J),$ is similar to the notation for the matrix representing  a Poincar\'{e} transformation. For a Poincar\'{e} transformation that combines a homogeneous Lorentz transformation $\Lambda$ with a translation through displacement $b,$ the matrix representing the transformation is written $D(\Lambda,b),$ with 
\begin{equation} \label{Poincare} D(\Lambda,b)D(\bar{\Lambda},\bar{b}) = D(\Lambda \bar{\Lambda},\Lambda \bar{b} +b) \quad .\end{equation}

There are two sets of momentum matrices $D(P)$ that satisfy the commutation rules of the Poincar\'{e} algebra. Nonzero momentum matrices occur for spins $(A,B) \oplus (C,D)$ with
\begin{equation} \label{spintype} \mid A-C\mid = \mid B-D\mid = \frac{1}{2} \quad .  \end{equation}
The momentum matrices $P_{\mu}$ are triangular in this representation,
\begin{equation} \label{PP4} D(P_{\mu}) = \pmatrix{0 && (P_{\mu 12})_{ab,cd} \cr 0 && 0}    \hspace{1cm} {\mathrm{or}} \hspace{1cm}   D(P_{\mu}) = \pmatrix{0 && 0 \cr (P_{\mu 21})_{cd,ab} && 0} .  \end{equation}
where $P_{\mu 12 }$ and $P_{\mu 21 }$ are off-diagonal blocks whose components are given in the Appendix. In this paper the 12 representation is discussed. The 21 representation gives similar results.

The double index $ab$ indicates the $z$-components of angular momentum, see (\ref{JABz}) and (\ref{KABz}); index $a$ can take $2A+1$ values $\{-A,-A+1, \ldots, A-1,A\}$ and $b$ can have any of the $2B+1$ values from $-B$ to $B.$ The double index $ab$ can be replaced by a single index $i,$
\begin{equation} \label{indexi1} i = (A+a)(2B+1) +B+b+1  \quad .\end{equation}
The $cd$ indices begin where the $ab$ indices stop, so the single index $i$ for the double index  $cd$ is given by
\begin{equation} \label{indexi2} i = (2A+1)(2B+1) +(C+c)(2D+1) +D+d+1 \quad .\end{equation}
The single index $i$ runs from $1$ to $d,$ the dimension of the representation, 
\begin{equation} \label{dim} d = (2A+1)(2B+1) +(2C+1)(2D+1)\end{equation}
and is naturally split into the $AB$ block consisting of $1$ through $(2A+1)(2B+1)$ and the $CD$ block of $(2A+1)(2B+1) +1$ through $d.$

\section{Little Group} \label{LG}

The little group of the light-like vector $k^{\mu}$ = $\{0,0,k^t,k^t\}$ consists of homogeneous Lorentz transformations $W$ that leave $k^{\mu}$ invariant. The generators of the little group,$^{ \cite{Wigner2}-\cite{W2}}$ $J$ = $J_z,$ $L_1$ = $J_y +K_x,$ and $L_2$ = $ -J_x + K_y,$ obey the commutation rules of the Euclidean group $E_2$ of motions in the plane,
\begin{equation} \label{JAB} [J,L_1] =  iL_2; \quad  [J,L_2] = - iL_1; \quad [L_1, L_2] = 0  \quad .\end{equation}
The same commutation rules are obeyed by $J_z,$ $P_x,$ and $P_y.$ Thus $L_1$ and $L_2$ generate the translations of $E_2$. These must be thought of as translations in some abstract $\alpha \beta$-plane, not the $xy$-plane. Thus any little group transformation $W$ can be written in the form
\begin{equation} \label{W} W(\alpha, \beta, \theta) = S(\alpha, \beta ) R(\theta) = \exp{i(\alpha L_1 + \beta L_2)} \exp{(i \theta J)} \quad , \end{equation}
where, following Ref. \cite{W2}, $R(\theta)$ denotes a rotation through an angle $\theta$ about the origin in the  $\alpha \beta$-plane and $S(\alpha, \beta )$ denotes an $\alpha \beta$-translation.

Since $D(R,0)$ is diagonal in the covariant non-unitary representation used here, see Appendix A, it is convenient to take the eigenvectors of $D(R,0)$ as a basis for the $d$-dimensional vectors. The $j$th component of the $i$th eigenvector can be taken to be
\begin{equation} \label{vi} (v_i)_j = \delta_{ij} \quad {\mathrm{or}} \quad (v_{ab})_{\bar{a} \bar{b}} = \delta_{a \bar{a}} \delta_{b \bar{b}}  \quad ,\end{equation}
where the second expression is for the $AB$ block in double index notation. The eigenvectors are the columns of the $d$-dimensional unit matrix. The eigenvalues can be obtained from (\ref{JABz}),  
\begin{equation} \label{lambdai} D(R(\theta),0)\pmatrix{ v_{ab} \cr v_{cd}} = \pmatrix{ e^{i(a+b)\theta} v_{ab} \cr e^{i(c+d)\theta} v_{cd} }  \quad , \end{equation}
where $v_{ab}$ is in the $AB$ block and $v_{cd}$ is in the $CD$ block. A general vector $u_i$ can be written in block notation,
\begin{equation} \label{c1} u_i = \sum_{ab} u_{ab} v_{ab} + \sum_{cd} u_{cd} v_{cd} = \pmatrix{u_{ab} \cr u_{cd} }  \quad ,  \end{equation}
where the notation indicates the natural separation of the indices into the $AB$ and $CD$ blocks.

Some vectors have special properties that can be deduced from the rows and columns of the matrix $\alpha L_1 + \beta L_2, $ which generates the $E_2$ translations. By (\ref{JAB+}), (\ref{KAB+}), and (\ref{JAB}) the components of $\alpha L_1 + \beta L_2$ are given by
\begin{equation} \label{alphaAbetaB} D(\alpha L_1 + \beta L_2)_{ab, \bar{a}\bar{b}}, = -i(\alpha - i \beta) r^{A}_{\bar{a}} \delta_{a ,\bar{a}+1}\delta_{b ,\bar{b}} + i(\alpha + i \beta) r^{B}_{-\bar{b}} \delta_{a ,\bar{a}}\delta_{b ,\bar{b}-1} \end{equation}
$$ D(\alpha L_1 + \beta L_2)_{cd, \bar{c}\bar{d}}, = -i(\alpha - i \beta) r^{C}_{\bar{c}} \delta_{c ,\bar{c}+1}\delta_{d ,\bar{d}} + i(\alpha + i \beta) r^{D}_{-\bar{d}} \delta_{c ,\bar{c}}\delta_{d ,\bar{d}-1} \quad .$$
Note that $\alpha L_1 + \beta L_2 $ is a block-diagonal matrix; the off-diagonal blocks 12 and 21 are zero.

Inspection of the delta functions in (\ref{alphaAbetaB}) shows that $\alpha L_1 + \beta L_2$ has two zero columns: the $(\bar{a},\bar{b})$ = $(A,-B)$ and the $(\bar{c},\bar{d})$ = $(C,-D)$ columns. This implies that $v_{A,-B}$ and $v_{C,-D}$ are invariant under $E_2$-translations and, hence, are block-by-block eigenvectors of $D(W(\alpha,\beta,\theta))$ with their $D(R(\theta),0)$-eigenvalues,
\begin{equation} \label{SWR1}   \pmatrix{W^{(A,B)} & 0 \cr 0 & W^{(C,D)} } \pmatrix{v_{A,-B} \cr v_{C,-D}} = \pmatrix{e^{ i(A-B)\theta} v_{A,-B} \cr e^{i(C-D)\theta } v_{C,-D} } \quad .  \end{equation}
The vectors $v_{A,-B}$ and $v_{C,-D}$ are called $k^{\mu}$-like vectors since $v_{A,-B} \cong$ $k^{\mu}$ in the ordinary 4-vector representation of the Lorentz group for spin $(A,B)$ = $(1/2,1/2).$ (The 4-vector representation differs from the representation here for spin (1/2,1/2) by a similarity transformation.) Equation (\ref{SWR1}) is the generalization to other spins of the statement that the transformation $W$ leaves $k^{\mu}$ invariant.

Adjacent columns in (\ref{alphaAbetaB}) have just one component, columns $(A,-B+1)$ and $(A-1,-B)$ have a nonzero component at row $(A,-B)$ and columns $(C,-D+1)$ and $(C-1,-D)$ have a nonzero component at row $(C,-D).$ The deltas in (\ref{alphaAbetaB}) also show that the row $(a,b)$ =  $(A,-B)$ has just the two nonzero components at columns $(A,-B+1)$ and $(A-1,-B)$ and the row $(c,d)$ = $(C,-D)$ has just the two nonzero components at columns $(C,-D+1)$ and $(C-1,-D).$ The vectors $v_{A,-B+1}$ and $v_{A-1,-B}$ play a role in the work here with the 12 momentum representation, while the vectors $v_{C,-D+1}$ and $v_{C-1,-D}$ would be used with the 21 momentum representation.

By (\ref{alphaAbetaB}) the adjacent vectors, $v_{A-1,-B}$ and $v_{A,-B+1}$ obey connection-like transformation rules,  
\begin{equation} \label{W1}   D(W,0) \cdot \pmatrix{v_{A-1,-B} \cr 0} = \pmatrix{e^{ i(A-B-1)\theta} v_{A-1,-B} \cr 0}  + (\alpha - i \beta) e^{-i\theta}\sqrt{2A} \pmatrix{e^{ i(A-B)\theta} v_{A,-B} \cr 0}  \end{equation}
\begin{equation} \label{W2}   D(W,0) \cdot \pmatrix{ v_{A,-B+1} \cr 0} = \pmatrix{ e^{ i(A-B+1)\theta} v_{A,-B+1} \cr 0}  -  (\alpha + i \beta) e^{i\theta}\sqrt{2B} \pmatrix{e^{ i(A-B)\theta} v_{A,-B} \cr 0}   \quad . \end{equation}
As is well-known, the appearance of the $k^{\mu}$-like vector $v_{A,-B}$ indicates a gauge transformation in the photon interpretation of the spin $(1/2,1/2)$ case. Hence, from the little group, gauge terms appear with $E_2$-translations through $(\alpha,\beta)$ and are absent without such translations. 

Actual spacetime translations are part of the Poincar\'{e} group, though not part of the little group. Thus the question becomes: under what circumstances do spacetime translations produce gauge terms?

\section{Spacetime Translations } \label{ST}

Translations are generated by momentum matrices in one of the two forms $P_{12}$ or $P_{21}$ in (\ref{PP4}). Thus there are two different representations, one with the 12 block of the $P^{\mu}$ nonzero and one with the 21 block nonzero. As mentioned previously, only the 12-representation is discussed in this paper. The treatment of the 21-representation is similar.

Translation through a spacetime displacement of $\epsilon^{\mu}$ is represented by the matrix
\begin{equation} \label{T} D(1,\epsilon^{\mu})  = \exp{(-i \eta_{\mu \nu} \epsilon^{\nu}P^{\mu})} =  \pmatrix{1 && -i \epsilon_{\mu}P^{\mu}_{12} \cr 0 && 1} \quad ,\end{equation}
where $\epsilon_{\mu}$ = $\eta_{\mu \nu} \epsilon^{\nu},$ the metric $\eta$ is diagonal, $\eta$ = diag$\{1,1,1,-1\},$ and the components of $P_{12}$ are given in the Appendix, (\ref{pp+12}) - (\ref{mmz21}). The higher powers of off-diagonal matrices $D(P^{\mu})$ vanish since
\begin{equation} \label{block} D(P^{\mu})D(P^{\nu}) =  \pmatrix{0 && P^{\mu}_{12} \cr 0 && 0} \pmatrix{0 && P^{\nu}_{12} \cr 0 && 0} = \pmatrix{0 && 0 \cr 0 && 0}  \quad . \end{equation}
Applying the translation to a vector $u$ yields
\begin{equation} \label{T1} D(1,\epsilon^{\mu})\pmatrix{u_{ab} \cr u_{cd}}  = \pmatrix{1 && -i\epsilon_{\mu} P^{\mu}_{12} \cr 0 && 1}\pmatrix{u_{ab} \cr u_{cd}} = \pmatrix{u_{ab} -i\epsilon_{\mu} {P^{\mu}_{12}}_{ab,\bar{c} \bar{d}} u_{\bar{c} \bar{d}} \cr u_{cd}} \quad . \end{equation}

Thus all $AB$ block vectors, $u_{cd}$ = 0, are invariant under translation with $D(1,\epsilon^{\mu}).$ Also, applying $D(1,\epsilon^{\mu})$ to the $CD$ block $k^{\mu}$-like vector $v_{C,-D}$ produces a connection-like transformation. Since the components of $P^{\mu}_{12}$ depend on spin type, there are four different results:
\pagebreak
\vspace{1cm}
\noindent Type 1: $A$ = $C+1/2$ and $B$ = $D+1/2$
 $$D(1,\epsilon^{\mu})v_{C,-D} = v_{C,-D} - i (\epsilon^{x} - i \epsilon^{y}) \frac{K^{12}}{\sqrt{2B}}  v_{A,-B+1} + i (\epsilon^{x} + i \epsilon^{y}) \frac{K^{12}}{\sqrt{2A}} v_{A-1,-B}     $$ 
\begin{equation} \label{Tv1} \quad \hspace{4cm} + i (\epsilon^{z}-\epsilon^{t})  \frac{K^{12}}{2\sqrt{AB}}v_{A-1,-B+1} + i  (\epsilon^{z}+\epsilon^{t}) K^{12}v_{A,-B}  \quad , \end{equation}
\noindent Type 2: $A$ = $C+1/2$ and $B$ = $D-1/2$
\begin{equation} \label{Tv2} D(1,\epsilon^{\mu})v_{C,-D} = v_{C,-D} - i (\epsilon^{x} - i \epsilon^{y}) K^{12} \sqrt{2D}  v_{A,-B} + i (\epsilon^{z}-\epsilon^{t})  K^{12} \sqrt{\frac{D}{A}} v_{A-1,-B}   \quad ,  \end{equation}
\noindent Type 3: $A$ = $C-1/2$ and $B$ = $D+1/2$
\begin{equation} \label{Tv3} D(1,\epsilon^{\mu})v_{C,-D} = v_{C,-D} - i (\epsilon^{x} + i \epsilon^{y}) K^{12} \sqrt{2C}  v_{A,-B} - i (\epsilon^{z}-\epsilon^{t})  K^{12} \sqrt{\frac{C}{B}} v_{A,-B+1}   \quad ,  \end{equation} 
\noindent Type 4: $A$ = $C-1/2$ and $B$ = $D-1/2$
\begin{equation} \label{Tv4}  D(1,\epsilon^{\mu})v_{C,-D} = v_{C,-D} - 2i (\epsilon^{z}-\epsilon^{t}) K^{12} \sqrt{CD}  v_{A,-B}  \quad .   \hspace{4cm}\end{equation} 

By (\ref{Tv1})-(\ref{Tv4}), the displacements can be chosen so that only the $k^{\mu}$-like vector $v_{A,-B}$ appears in the $AB$ block. For example, for spin type 1 and real
displacements $\epsilon^{\mu},$ the coefficients of $v_{A-1,-B},$ $v_{A,-B+1},$ and $v_{A-1,-B+1}$ vanish leaving only the $k^{\mu}$-like $v_{A,-B}$ term
\begin{equation} \label{Tv5} D(1,\epsilon^{\mu})\pmatrix{0 \cr v_{C,-D}} = \pmatrix{0 \cr v_{C,-D}}  +  2i\epsilon^{t} K^{12}\pmatrix{v_{A,-B} \cr 0} \quad {\mathrm{(Type}} \enspace {\mathrm{1)}} \quad , \end{equation}
for $\epsilon^{\mu}$ = $\{0,0,\epsilon^{t},\epsilon^{t}\}.$ The results for all four spin types are collected in Table 1. Note that for spin types 1, 2, and 3, the required displacements $\epsilon^{\mu}$ have a vanishing inner product with the standard momentum $k^{\mu},$ $\epsilon_{\mu}k^{\mu}$ = 0. It is interesting that light-like displacements along $k^{\mu}$ arise almost independently of the momentum $k^{\mu},$ the only association being the elimination of any vector $v_{ab}$ that is not $k^{\mu}$-like from the result of translating the $k^{\mu}$-like vector $v_{C,-D}.$

\begin{table}[tbp] \label{tx1} \caption{Displacements $\epsilon^{\mu}$ for each spin type. Only $k^{\mu}$-like gauge terms with $v_{A,-B}$ and $v_{C,-D}$ survive these displacements in  (\ref{Tv1})-(\ref{Tv4}). }  \begin{tabular}{|c||c|c|c|c|} \hline {Spin Type} & { 1} & 2  & 
3 & 4 \\ \hline \hline {Displacement} &{ {$ \{0,0,\epsilon^t,\epsilon^t\}$}} &  {$ \{\epsilon^x,\epsilon^y,\epsilon^t,\epsilon^t\}$} & {$   \{\epsilon^x,\epsilon^y,\epsilon^t,\epsilon^t \}$} & {$ \{\epsilon^x,\epsilon^y,\epsilon^z,\epsilon^t\}$}\\ \hline \end{tabular} \end{table}

\section{Little Group  \& Spacetime Translations} \label{LGST}

Now the gauge terms, i.e. the $k^{\mu}$-like contributions $v_{A,-B}$, produced by spacetime translations and little group translations can be combined. 
 Little group displacements provide gauge terms when applied to the adjacent vectors $v_{A-1,-B}$ and $v_{A,-B+1}$ and spacetime displacements give gauge terms when applied to the $k^{\mu}$-like vector $v_{C,-D}.$ A linear combination of these terms can be obtained by combining the $AB$ and $CD$ block vectors to make `polarization -like' vectors $e_{\pm}$ defined by
\begin{equation} \label{Ev1} e_{-} = \pmatrix{v_{A-1,-B} \cr c_1 v_{C,-D}} \quad {\mathrm{and}} \quad e_{+} = \pmatrix{v_{A,-B+1} \cr c_2 v_{C,-D}} \quad ,\end{equation} 
where $c_i$ are arbitrary coefficients. 

The matrix representing a spacetime translation through a displacement $\epsilon$ followed by a little group transformation $W$ has the form
\begin{equation} \label{WT} D(W(\alpha,\beta,\theta),0)D(1,\epsilon^{\mu}) = \pmatrix{W^{(A,B)} & -i \epsilon_{\mu}W^{(A,B)}P^{\mu}_{12}\cr 0 & W^{(C,D)}}  \quad . \end{equation} 
Note that, in this representation, $D(W)$ components appear in the diagonal blocks while the displacement effects are produced by an off-diagonal block. Applying $D(W,0)D(1,\epsilon)$ to the vectors $e_{-}$ and $e_{+}$ gives different results for each spin type,

\noindent Type 1: $A$ = $C+1/2;$ $B$ = $D+1/2,$ for $\epsilon^{\mu}$ = $\{0,0,\epsilon^{t},\epsilon^{t}\},$ i.e. $\epsilon^{\mu} \propto k^{\mu},$
 \begin{equation} \label{WTv1} D(W,0)D(1,\epsilon)e_{-} = \lambda e_{-} + [ 2i c_1 K^{12}\epsilon^{t}+(\alpha-i\beta)e^{-i\theta}\sqrt{2A} ]\pmatrix{e^{-i(A-B)\theta}v_{A,-B} \cr 0} \end{equation} 
 \begin{equation} \label{WTv2} D(W,0)D(1,\epsilon)e_{+} = \lambda e_{+} + [ 2i c_1 K^{12}\epsilon^{t}-(\alpha+i\beta)e^{i\theta}\sqrt{2B} ]\pmatrix{e^{i(A-B)\theta}v_{A,-B} \cr 0}    \end{equation} 
\noindent Type 2: $A$ = $C+1/2;$ $B$ = $D-1/2,$ for $\epsilon^{\mu}$ = $\{\epsilon^{x},\epsilon^{y},\epsilon^{t},\epsilon^{t}\},$ i.e. $\epsilon^{z}=\epsilon^{t},$
 \begin{equation} \label{WTv3} D(W,0)D(1,\epsilon)e_{-} = \lambda e_{-} - [i c_1 K^{12}\sqrt{2D}(\epsilon^{x}-i\epsilon^{y})-(\alpha-i\beta)e^{-i\theta}\sqrt{2A} ]\pmatrix{e^{-i(A-B)\theta}v_{A,-B} \cr 0}    \end{equation} 
 \begin{equation} \label{WTv4} D(W,0)D(1,\epsilon)e_{+} = \lambda e_{+} - [ i c_1 K^{12}\sqrt{2D}(\epsilon^{x}-i\epsilon^{y})+(\alpha+i\beta)e^{i\theta}\sqrt{2B} ]\pmatrix{e^{i(A-B)\theta}v_{A,-B} \cr 0}    \end{equation} 
\noindent Type 3: $A$ = $C-1/2;$ $B$ = $D+1/2,$ for $\epsilon^{\mu}$ = $\{\epsilon^{x},\epsilon^{y},\epsilon^{t},\epsilon^{t}\},$ i.e. $\epsilon^{z}=\epsilon^{t},$
 \begin{equation} \label{WTv5} D(W,0)D(1,\epsilon)e_{-} = \lambda e_{-} - [i c_1 K^{12}\sqrt{2C}(\epsilon^{x}+i\epsilon^{y})-(\alpha-i\beta)e^{-i\theta}\sqrt{2A} ]\pmatrix{e^{-i(A-B)\theta}v_{A,-B} \cr 0}    \end{equation} 
 \begin{equation} \label{WTv6} D(W,0)D(1,\epsilon)e_{+} = \lambda e_{+} - [ i c_1 K^{12}\sqrt{2C}(\epsilon^{x}+i\epsilon^{y})+(\alpha+i\beta)e^{i\theta}\sqrt{2B} ]\pmatrix{e^{i(A-B)\theta}v_{A,-B} \cr 0}   \end{equation} 
\noindent Type 4: $A$ = $C-1/2$ and $B$ = $D-1/2$ for $\epsilon^{\mu}$ = $\{\epsilon^{x},\epsilon^{y},\epsilon^{z},\epsilon^{t}\},$ i.e. $\forall \enskip \epsilon^{\mu},$
 \begin{equation} \label{WTv7} D(W,0)D(1,\epsilon)e_{-} = \lambda e_{-} - [2i c_1 K^{12}\sqrt{CD}(\epsilon^{z}-\epsilon^{t})-(\alpha-i\beta)e^{-i\theta}\sqrt{2A} ]\pmatrix{e^{-i(A-B)\theta}v_{A,-B} \cr 0}    \end{equation} 
 \begin{equation} \label{WTv8} D(W,0)D(1,\epsilon)e_{+} = \lambda e_{+} - [ 2i c_1 K^{12}\sqrt{CD}(\epsilon^{z}-\epsilon^{t})+(\alpha+i\beta)e^{i\theta}\sqrt{2B} ]\pmatrix{e^{i(A-B)\theta}v_{A,-B} \cr 0}    \end{equation} 
where 
\begin{equation} \label{Du5a}
   \lambda e_{-} = \pmatrix{e^{i(A-B-1)\theta}v_{A-1,-B} \cr e^{i(C-D)\theta}c_1 v_{C,-D}} \quad {\mathrm{and}} \quad \lambda e_{+} = \pmatrix{e^{i(A-B+1)\theta}v_{A,-B+1} \cr e^{i(C-D)\theta}c_1 v_{C,-D}} \quad .
\end{equation}
Note that there is a different eigenvalue in each block, not one overall eigenvalue for both blocks. Call them {\it{blockwise}} eigenvectors. 

The polarization-like vectors $e_{-}$ and $e_{+}$ are blockwise eigenvectors of $D(W,0)D(1,\epsilon)$ if the gauge terms in (\ref{WTv1})-(\ref{WTv8}) cancel. As discussed in the introduction, determining $\alpha$ and $\beta$ eliminates some unwanted degrees of freedom.$^{ \cite{VDBij}- \cite{W5}}$  Therefore, allowing little group rotation angle $\theta$ and spacetime displacement $\epsilon^{\mu}$ to vary freely, one finds formulas for little group displacements $\alpha$ and $\beta$ that depend on spin type,

\noindent Type 1: $A$ = $C+1/2;$ $B$ = $D+1/2,$ for $\epsilon^{\mu}$ = $\{0,0,\epsilon^{t},\epsilon^{t}\},$ i.e. $\epsilon^{\mu} \propto k^{\mu},$
 \begin{equation} \label{WTv1a} D(\bar{W},0)D(1,\epsilon)e_{-} = \lambda e_{-} \quad {\mathrm{for}} \quad \alpha-i\beta =  -i \frac{\sqrt{2} c_1 K^{12}}{\sqrt{A}}\epsilon^{t} e^{i\theta}   \quad ,   \end{equation} 
 \begin{equation} \label{WTv2a} D(\bar{W},0)D(1,\epsilon)e_{+} = \lambda e_{+}  \quad {\mathrm{for}} \quad \alpha+i\beta = + i \frac{\sqrt{2}c_1 K^{12}}{\sqrt{B}}\epsilon^{t} e^{-i\theta}   \quad ,   \end{equation} 
\noindent Type 2: $A$ = $C+1/2;$ $B$ = $D-1/2,$ for $\epsilon^{\mu}$ = $\{\epsilon^{x},\epsilon^{y},\epsilon^{t},\epsilon^{t}\},$ i.e. $\epsilon^{z}=\epsilon^{t},$
 \begin{equation} \label{WTv3a} D(\bar{W},0)D(1,\epsilon)e_{-} = \lambda e_{-} \quad {\mathrm{for}} \quad \alpha-i\beta =  +i c_1 K^{12}\sqrt{\frac{D}{A}}(\epsilon^{x}-i\epsilon^{y}) e^{+i\theta}   \quad ,  \end{equation} 
 \begin{equation} \label{WTv4a} D(\bar{W},0)D(1,\epsilon)e_{+}  = \lambda e_{+} \quad {\mathrm{for}} \quad \alpha+i\beta =  -i c_1 K^{12}\sqrt{\frac{D}{B}}(\epsilon^{x}-i\epsilon^{y}) e^{-i\theta}  \quad , \end{equation} 
\noindent Type 3: $A$ = $C-1/2;$ $B$ = $D+1/2,$ for $\epsilon^{\mu}$ = $\{\epsilon^{x},\epsilon^{y},\epsilon^{t},\epsilon^{t}\},$ i.e. $\epsilon^{z}=\epsilon^{t},$
 \begin{equation} \label{WTv5a} D(\bar{W},0)D(1,\epsilon)e_{-} = \lambda e_{-} \quad {\mathrm{for}} \quad \alpha-i\beta =  +i c_1 K^{12}\sqrt{\frac{C}{A}}(\epsilon^{x}+i\epsilon^{y}) e^{+i\theta}    \quad ,  \end{equation} 
 \begin{equation} \label{WTv6a} D(\bar{W},0)D(1,\epsilon)e_{+} = \lambda e_{+} \quad {\mathrm{for}} \quad \alpha+i\beta =  -i c_1 K^{12}\sqrt{\frac{C}{B}}(\epsilon^{x}+i\epsilon^{y}) e^{-i\theta}  \quad , \end{equation} 
\noindent Type 4: $A$ = $C-1/2$ and $B$ = $D-1/2$ for $\epsilon^{\mu}$ = $\{\epsilon^{x},\epsilon^{y},\epsilon^{z},\epsilon^{t}\},$ i.e. $\forall \enskip \epsilon^{\mu},$
 \begin{equation} \label{WTv7a} D(\bar{W},0)D(1,\epsilon)e_{-} = \lambda e_{-} \quad {\mathrm{for}} \quad \alpha-i\beta = +i c_1 K^{12}\sqrt{\frac{2CD}{A}}(\epsilon^{z}-\epsilon^{t}) e^{+i\theta}    \quad ,  \end{equation} 
 \begin{equation} \label{WTv8a} D(\bar{W},0)D(1,\epsilon)e_{+} = \lambda e_{+} \quad {\mathrm{for}} \quad \alpha+i\beta =  -i c_1 K^{12}\sqrt{\frac{2CD}{B}}(\epsilon^{z}-\epsilon^{t}) e^{-i\theta}   \quad ,  \end{equation}
where $\bar{W}$ indicates the function 
\begin{equation} \label{W0}
\bar{W}(\epsilon^{\mu},\theta) = W(\alpha(\epsilon^{\mu},\theta),\beta(\epsilon^{\mu},\theta),\theta)  
\end{equation}
and the functions $\alpha$ and $\beta$ are as given in (\ref{WTv1a})-(\ref{WTv8a}).
Thus the gauge terms can cancel and the polarization-like vectors $e_{\pm}$ are then block-by-block eigenvectors of the little group and translation combination $D(\bar{W},0)D(1,\epsilon).$

The determination of the parameters $\alpha, \beta$ reinforces the restrictions needed to produce the unitary transformation of annihilation fields with (\ref{Da}). In the derivation of (\ref{Da}) it is assumed that the degrees of freedom represented by $\alpha, \beta$ do not distinguish different particle states. If the parameters $\alpha, \beta$ are allowed to be general real numbers with nontrivial generators $L_1$ and $L_2,$ then the resulting infinity of particle states would absorb infinite energy to achieve thermal equilibrium. Thus it is said that specific heats would become infinite.$^{ \cite{VDBij}- \cite{W5}}$ So, to avoid this, the unitary generators $L_1$ and $L_2$ are made to vanish.$^{\cite{W5}}$ The parameter $\theta$ is the helicity operator for photons and its generator $J$ is therefore kept.

The determination of $\alpha, \beta$ in terms of rotation angle $\theta$ and displacement $\epsilon^{\mu}$ does much the same for the covariant representation. For the covariant representation, the degrees of freedom for $\alpha, \beta$ are frozen because the parameters are not allowed to vary. Perhaps canceling gauge terms explains why these degrees of freedom are frozen.

\section{Covariant and Canonical Fields } \label{CCF}

The gauge terms from little group translations appear in standard constructions of a covariant vector field $\psi$ as a sum over particle states. As just seen one can cancel the gauge terms from little group translations with gauge terms from spacetime translations for those spin types that have nonzero matrix momentum generators. Thus, it may be that gauge invariance need not be invoked for those spin types that have nonzero momentum matrices. However, as seen below, gauge invariance is required for all but some special momenta and transformations.

The construction of covariant fields $\psi$ in terms of particles states $a$ is a well known procedure. For setting the notation and for convenience a short reminder of the formalism follows.$^{ \cite{W3}, \cite{W4}}$ 

The covariant vector field $\psi$ has both a continuous and a discrete dependence, the coordinate $x^{\mu}$ and spin index $l$, $\{x^{\mu},l\}$ whereas the canonical vector field $a$ has momentum and helicity indices, $\{p^{\mu},\sigma\}$. Thus the constructing a field $\psi$ as a linear combination of fields $a$ takes the form 
\begin{equation} \label{psiLP}
\psi_{l}(x) = \sum_{\sigma} \int d^3 p \enspace u_{l}(x,{\mathbf{p}},\sigma) a({\mathbf{p}},\sigma)  \quad ,
\end{equation}
where the coefficients $u$ are called intertwiners and $\mathbf{p}$ indicates the space components of the momentum, $\{p^{x},$$p^{y},$$p^{z}\}.$ Note that the single index notation, here $l,$ is used rather than the double index notation,  $ab$ or $cd,$ which is found in many of the previous expressions. Only positive energy fields are treated here. The canonical vectors $a$ are also known as annihilation operators. The discussion for negative energy and creation operators can be transcribed from the discussion here. 

A spacetime transformation $\{\Lambda , b\}$ results from a Lorentz transformation $\Lambda$ followed by a translation through displacement 4-vector $b$ to make a Poincar\'{e} transformation. The construction transforms according to  
\begin{equation} \label{Dpsi1} 
U(\Lambda,b) \psi_{l}(x) {U}^{-1}(\Lambda,b) = \sum_{\sigma} \int d^3 p \enspace U(\Lambda,b) u_{l}(x,{\mathbf{p}},\sigma){U}^{-1}(\Lambda,b) U(\Lambda,b)  a({\mathbf{p}},\sigma) {U}^{-1}(\Lambda,b) \quad ,
\end{equation}
where the intertwiners are allowed to change under the transformation.
The covariant field transforms as
\begin{equation} \label{Dpsi2}
U(\Lambda,b) \psi_{l}(x) {U}^{-1}(\Lambda,b) = D^{-1}_{l \bar{l}}(\Lambda,b)  \psi_{\bar{l}}(\Lambda x + b)
\end{equation}
and the canonical vector field transforms as
\begin{equation} \label{Da}
U(\Lambda,b) a({\mathbf{p}},\sigma) {U}^{-1}(\Lambda,b) = \sqrt{\frac{(\Lambda p)^t}{p^t}} e^{-i \sigma \theta(p,\Lambda)} e^{i \Lambda p \cdot b}  a({\mathbf{\Lambda p}},\sigma) \quad ,
\end{equation}
where $\theta (p,\Lambda)$ is determined by writing the little group transformation 
\begin{equation} \label{WLp}
W(p, \Lambda) = L^{-1}(\Lambda p)\Lambda L(p) = S(\alpha(p, \Lambda),\beta(p, \Lambda))R(\theta(p, \Lambda))
\end{equation}
in the form $W$ = $S(\alpha,\beta)R(\theta)$ and $L(p) $ is a standard transformation taking the standard 4-vector $k^{\mu}$ to $p,$ first boosting $k^{\mu}$ in the $z$ direction so that $k^0 \rightarrow$ $p^0,$ then rotating the $z$-axis to the direction of $\mathbf{p},$ for details see Weinberg.$^{\cite{W5}}$ While the $\theta$ for little group rotations is present, there is no dependence in (\ref{Da}) on little group translations $\alpha,\beta.$ The conventional derivation of (\ref{Da}) eliminates these unobserved degrees of freedom by making the generators of little group translations vanish.$^{ \cite{VDBij}, \cite{W5}}$ 

Indicate the transformed intertwiners with a prime, 
\begin{equation} \label{Du1}
U(\Lambda,b) u_{l}(x,{\mathbf{p}},\sigma) {U}^{-1}(\Lambda,b) = u^{\prime}_{l}(x,{\mathbf{p}},\sigma,\Lambda,b) \quad .
\end{equation}
By (\ref{Dpsi2}) and (\ref{Da}), the intertwiners must transform as
\begin{equation} \label{Du2}
   D_{l \bar{l}}(\Lambda,b) u^{\prime}_{\bar{l}}(x,{\mathbf{p}},\sigma,\Lambda,b) = \sqrt{\frac{ (\Lambda p)^t}{p^t}} e^{i \sigma \theta(p,\Lambda)} e^{-i \Lambda p \cdot b} u_{l}(\Lambda x + b,{\mathbf{\Lambda p}},\sigma) \quad .
\end{equation}
The coordinate $x$ and translation $b$ can be taken care of by requiring that
\begin{equation} \label{Du2a}
  u^{\prime}_{l}(x,{\mathbf{p}},\sigma,\Lambda,b) = e^{i p \cdot x} D_{l \bar{l}}(1,x)  u^{\prime}_{\bar{l}}(0,{\mathbf{p}},\sigma,\Lambda,b) 
\end{equation}
and 
\begin{equation} \label{Du3}
  u_{l}(\Lambda x + b,{\mathbf{\Lambda p}},\sigma) = e^{i \Lambda p \cdot (\Lambda x + b)} D_{l \bar{l}}(1,\Lambda x + b)  u_{l}(0,{\mathbf{\Lambda p}},\sigma) \quad .
\end{equation}
Then (\ref{Du2}) becomes
\begin{equation} \label{Du2c}
   D_{l \bar{l}}(\Lambda,0) u^{\prime}_{\bar{l}}(0,{\mathbf{p}},\sigma,\Lambda,b) = \sqrt{\frac{ (\Lambda p)^t}{p^t}} e^{i \sigma \theta(p,\Lambda)}  u_{l}(0,{\mathbf{\Lambda p}},\sigma) \quad .
\end{equation}
For $\Lambda$ = $W$ and $p$ = $k,$ one finds that
\begin{equation} \label{Du4a}
   D_{l \bar{l}}(W,0) u^{\prime}_{l}(0,{\mathbf{k}},\sigma,W,b) = e^{i \sigma \theta(k,W)} u_{l}(0,{\mathbf{k}},\sigma) \quad .
\end{equation}
 As discussed in the introduction, the goal is to obtain invariant intertwiners, $u^{\prime}_{l}(x$,${\mathbf{p}}$,$\sigma$,$\Lambda$,$b)$ $ \rightarrow$ $ u_{l}(x,{\mathbf{p}},\sigma),$ which would imply, by (\ref{Du4a}), that suitable invariant intertwiners could be found if there were any  eigenvectors of $D(W,0).$ 

By (\ref{SWR1}), the $k^{\mu}$-like vectors $v_{A,-B}$ and $v_{C,-D}$ are blockwise eigenvectors of $D(W,0)$ and could be the basis for massless particle fields when the eigenvalues of the two blocks are equal,
\begin{equation} \label{sigma0}
   \sigma_0 = A-B = C-D  \quad .
\end{equation}
By (\ref{spintype}), this implies that the spin $(A,B) \oplus (C,D)$ is restricted to $A-C$ = $B-D$ = $\pm 1/2,$ i.e. spin type 1 or 4.
Therefore, for spin types 1 and 4, define
\begin{equation} \label{Ev2} e_{0} = \pmatrix{v_{A,-B} \cr c_1 v_{C,-D}}  \quad ,\end{equation} 
where $c_1$ is constant. Then (\ref{Du4a}) is satisfied when the transformed intertwiner is the same as the intertwiner,
\begin{equation} \label{E0a} u^{\prime}_{l}(0,{\mathbf{k}},\sigma_{0},W,b) = u(0,{\mathbf{k}},\sigma_{0}) = e_{0} \quad ,\end{equation} 
where $\sigma_0 = A-B.$

Han et al. (\cite{Han}) choose a $k^{\mu}$-like vector for neutrinos to avoid gauge terms. But the helicity for spin $(A,B)$ = $(1/2,1/2)$ would be $A-B$ = 0, and that is not the helicity of a photon. Matrix representations of translations could be used with $k^{\mu}$-like vectors and such ideas may be discussed elsewhere, but they would defocus the discussion here.  

The gauge terms in (\ref{W1}) and (\ref{W2})  prevent $v_{A-1,-B}$ and $v_{A,-B+1}$ from being the invariant intertwiners. But, with spacetime translations to cancel the gauge terms, the vectors $e_{\pm}$ are blockwise eigenvectors of $D(\bar{W},0)D(1,\epsilon),$ (\ref{WTv1a})-(\ref{WTv8a}). These are eigenvectors when the eigenvalues for the two blocks coincide,
\begin{equation} \label{sigmapm1}
   \sigma_{\pm} = A-B \pm 1 = C-D  \quad .
\end{equation}
In order to have nontrivial translation matrices, the spin $(A,B) \oplus (C,D)$ is restricted. thus, by (\ref{spintype}) and (\ref{sigmapm1}), $\sigma_{-}$ = $A-B-1$ occurs only when $A-C$ = $-(B-D)$ = $+1/2$ which is spin type 2 and $\sigma_{+}$ = $A-B+1$ occurs only when $A-C$ = $-(B-D)$ = $-1/2$ which is spin type 3.

Thus there are  eigenvectors of $D(\bar{W},0)D(1,\epsilon)$, but not $D(W,0)$ itself, and one can choose the transformed intertwiners to be
\begin{equation} \label{Du5b}
   u^{\prime}_{l}(0,{\mathbf{k}},\sigma_{\pm},\bar{W},b) = D(1,\epsilon^{\mu})u_{m}(0,{\mathbf{k}},\sigma_{\pm})
\end{equation}
with
\begin{equation} \label{Du5c}
   u(0,{\mathbf{k}},\sigma_{\pm})=e_{\pm}   \quad .
\end{equation}
The choices (\ref{Du5b}) and (\ref{Du5c}) satisfy equation (\ref{Du4a}) with $W \rightarrow$ $\bar{W}.$ The $u^{\prime}_{l}(0,{\mathbf{k}},\sigma,W,b)$ and $u_{l}(0,{\mathbf{k}},\sigma)$  for all other $(A,B)$ helicities, $\sigma \neq$ $\sigma_{\pm},$ are set to zero. (As just noted above, one could make an exception for $\sigma_{0}$ =$A-B.$) 

For general momenta $p$ and coordinates $x$, the intertwiner $u_{l}(x,{\mathbf{p}},\sigma)$ and the transformed intertwiner $u^{\prime}_{l}(x,{\mathbf{p}},\sigma,\Lambda,b)$ can be obtained by standard procedures from $u_{l}(0,{\mathbf{k}},\sigma)$ and $u^{\prime}_{l}(0,{\mathbf{k}},\sigma,W,b)$. One finds that (\ref{Du2}) is satisfied for the appropriate spin type by
\begin{equation} \label{up-1}
   u^{\prime}_{l}(x,{\mathbf{p}},\sigma_{0},\Lambda,b) = u_{l}(x,{\mathbf{p}},\sigma_{0}) = \sqrt{\frac{k^0}{p^0}} e^{i p \cdot x} D_{lm}(L(p) ,x) {e_{0}}_{m} \quad ,
\end{equation}
\begin{equation} \label{up0}
   u^{\prime}_{l}(x,{\mathbf{p}},\sigma_{\pm},\Lambda,b) = \sqrt{\frac{k^0}{p^0}} e^{i p \cdot x} D_{l m}( L(p)\tilde{S}^{-1}(p, \Lambda)  ,x){e_{\pm}}_{m}
\end{equation}
\begin{equation} \label{up1}
   u_{l}(x,{\mathbf{p}},\sigma_{\pm}) = \sqrt{\frac{k^0}{p^0}} e^{i p \cdot x} D_{lm}(L(p) ,x) {e_{\pm}}_{m} \quad ,
\end{equation}
where helicity $\sigma_{0}$ is for spin types 1 and 4, $\sigma_{-}$ is for spin type 2, $\sigma_{+}$ works with spin type 3, and
\begin{equation} \label{up2}
   \tilde{S}(p, \Lambda) = R^{-1}(\theta(p, \Lambda))S(\alpha(p, \Lambda),\beta(p, \Lambda))R(\theta(p, \Lambda))
\end{equation}
is similar to the translation $S(\alpha(p, \Lambda),\beta(p, \Lambda))$ of the little group transformation $W(p, \Lambda),$ (\ref{WLp}). It is just this little group translation $\tilde{S}$ that distinguishes the transformed intertwiner $u^{\prime}_{l}(x,{\mathbf{p}},\sigma_{\pm},\Lambda,b)$ from the intertwiner $u_{l}(x,{\mathbf{p}},\sigma_{\pm}).$ All other intertwiners and transformed intertwiners are set to zero, so that there are three helicities $\sigma_0$ = $A-B$ and $\sigma_{\pm}$ = $A-B \pm 1.$ 

Thus the massless particle field for each spin type is determined,

\noindent Type 1: $A$ = $C + 1/2;$ $B$ = $D + 1/2,$ helicity $\sigma_{0}$ = $A-B,$
 \begin{equation} \label{psi3} \psi_{l}(x) = D_{lm}(1,x) \sum_{\sigma} \int d^3 p \enspace \sqrt{\frac{k^0}{p^0}} e^{i p \cdot x} D_{mn}(L(p) ,0) {e_{0}}_{n} a({\mathbf{p}},\sigma_{0})
   \quad ,   \end{equation} 
\noindent Type 2: $A$ = $C+1/2;$ $B$ = $D-1/2,$ helicity $\sigma_{-}$ = $A-B-1,$
 \begin{equation} \label{psi4} \psi_{l}(x) = D_{lm}(1,x) \sum_{\sigma} \int d^3 p \enspace \sqrt{\frac{k^0}{p^0}} e^{i p \cdot x} D_{mn}(L(p) ,0) {e_{-}}_{n} a({\mathbf{p}},\sigma_{-})    \quad ,  \end{equation} 
\noindent Type 3: $A$ = $C-1/2;$ $B$ = $D+1/2,$ helicity $\sigma_{+}$ = $A-B+1,$
 \begin{equation} \label{psi5} \psi_{l}(x) = D_{lm}(1,x) \sum_{\sigma} \int d^3 p \enspace \sqrt{\frac{k^0}{p^0}} e^{i p \cdot x} D_{mn}(L(p) ,0) {e_{+}}_{n} a({\mathbf{p}},\sigma_{+})    \quad ,  \end{equation} 
\noindent Type 4: $A$ = $C-1/2$ and $B$ = $D-1/2$ helicity $\sigma_{0}$ = $A-B,$
 \begin{equation} \label{psi6} \psi_{l}(x) = D_{lm}(1,x) \sum_{\sigma} \int d^3 p \enspace \sqrt{\frac{k^0}{p^0}} e^{i p \cdot x} D_{mn}(L(p) ,0)  {e_{0}}_{n} a({\mathbf{p}},\sigma_{0})    \quad ,  \end{equation} 
where it must be remembered that the intertwiners for helicity $\sigma_{\pm}$ transform by (\ref{up0}). Note that the translation matrix $D(1,x)$ factors out because it does not depend on the momentum $p$ or helicity $\sigma.$ This complicates the use of the gradient, $-i \partial_{\mu}$ as the momentum operator. However such situations occur often in gauge theory. 

As discussed in the introduction, intertwiner dependence on transformation means that the covariant non-unitary and canonical unitary representations of the massless class of the Poincar\'{e} group have an essential difference. This difference prevents the construction from being invariant under transformations, i.e. $\psi$ = $u a$ does not transform to $\psi^{\prime}$ = $u a^{\prime}.$

\section{Plane Waves, Equivalences, and Motion}   \label{PW}

By (\ref{Du3}) the intertwiners depend on coordinates as a plane wave, $u_{m}(x,{\mathbf{p}},\sigma) \propto $ $ \exp{(ip\cdot x)}$ when the translation matrices are trivial, i.e. $K^{12}$ = 0. It is reasonable to define the plane waves $ \exp{(ip\cdot x_1)}$ and $ \exp{(ip\cdot x_2)}$ to be equivalent if the difference $x_{2}-x_{1}$ obeys $p\cdot (x_{2}-x_{1})$ = 0. Equivalence is connected to motion; the motion of a plane wave is determined by the spacetime paths of constant phase $p \cdot x.$ These notions of motion and equivalent plane waves under coordinate translations can be extended to translations represented by matrices. 

Since the displacements $\epsilon^{\mu}$ for spin types 1, 2 and 3 obey $k \cdot \epsilon$ = 0, by (\ref{WTv1})-(\ref{WTv8}), the displacements take one plane wave to an equivalent one, $ \exp{(ik \cdot x) \rightarrow}$ $ \exp{(ik \cdot (x+\epsilon)}.$ Thus the motion is parallel to the momentum. 

For all four spin types, one may define the translated polarization-like vectors $D(1,\epsilon^{\mu})e_{\pm}$ to be {\it{equivalent}} to the vectors themselves, 
\begin{equation} \label{Du5f}
   D(1,\epsilon^{\mu})e_{\pm} \cong e_{\pm} \quad .
\end{equation}
 Then, by (\ref{Du2a}), (\ref{Du3}), (\ref{Du5b}) and (\ref{Du5c}) one finds that, 
\begin{equation} \label{Du5f1}
   u^{\prime}_{l}(x,{\mathbf{k}},\sigma_{\pm},\bar{W},b) \cong u_{m}(x,{\mathbf{k}},\sigma_{\pm}) \quad ,
\end{equation}
which means that the desired intertwiner invariance holds for any translation by $b$ and at least for one momentum, i.e. $k^{\mu},$ and at least one Lorentz transformation, i.e. $\bar{W}$. 

More generally, the equivalence can be defined based on (\ref{up0}) and (\ref{up1}). By (\ref{Poincare}), it follows that $D( L(p)\tilde{S}^{-1}  ,x) $ = $D( L(p),x)D( \tilde{S}^{-1},0) $, and intertwiner invariance holds when  
\begin{equation} \label{Du5g}
   D(\tilde{S}^{-1}(p, \Lambda),0)e_{\pm} \cong e_{\pm} \quad .
\end{equation}
Then, by (\ref{Du2a}) and (\ref{Du3}),
\begin{equation} \label{Du5g1}
   u^{\prime}_{l}(x,{\mathbf{p}},\sigma_{\pm},\Lambda,b) \cong u_{m}(x,{\mathbf{p}},\sigma_{\pm}) \quad .
\end{equation}
Note carefully that $\tilde{S}^{-1}(p, \Lambda)$ is a little group transformation and it produces gauge terms when applied to $e_{\pm},$ by (\ref{W1}) and (\ref{W2}). Therefore the equivalence (\ref{Du5g}) implies invariance under gauge transformations. 

Furthermore the properties of $e_{\pm}$ with $\bar{W}$ and $D(1,\epsilon)$ imply that (\ref{Du5f1}) is a special case of (\ref{Du5g1}). Thus, gauge invariance is not needed in the special case of $k^{\mu}$ with $\bar{W}$ where it can be replaced by invariance with certain spacetime translations. However the general case of any momentum $p$ and any Lorentz transformation $\Lambda$ requires gauge invariance.

The massless fields obtained with canceling the gauge terms for $p$ = $k$ and $\Lambda$ = $\bar{W}$ have only two definite helicities. The helicity can be either
\begin{equation} \label{sigma1} \sigma_{-} = A-B-1 \quad 
\end{equation}
for spin type 2 or 
\begin{equation} \label{sigma2}  \sigma_{+} = A-B+1     \quad ,
\end{equation}
for spin type 3.

There are no such massless scalar fields since spin $(A,B)$ = $(0,0)$ doesn't allow either of the adjacent vectors $v_{A-1,-B}$ or $v_{A,-B+1}.$ (The range of subscripts $ab$ collapses to just one, i.e. $ab$ = $00,$ so no vector has a subscript of $-1,0$ or $0,+1.$) For spin $(A,B)$ = $(0,1/2)$ there is just one helicity $\sigma_{+}$ = $1/2$ with $e_{+}$ and spin type 3, since there is no $e_{-}$ in this case. For spin $(A,B)$ = $(1/2,0)$ the helicity is $-1/2$ for $e_{-}$ with spin type 2.  For the photon, spin $(A,B)$ = $(1/2,1/2),$ there are the two expected helicities $\pm1,$  $\sigma_{-}$ = $-1$ with spin type 2 and $\sigma_{+}$ = $+1$ with spin type 3. Moreover the polarization-like vectors $e_{\pm}$ are the photon's usual transverse polarization vectors. But the helicity of the graviton, spin $(A,B)$ = $(1,1),$ would also be $\sigma_{\pm}$ = $\pm 1$ and that disagrees with conventional expectations that the helicity should be extreme, i.e. $\pm 2.$ Investigating the comparison with experiment entails further study.

The interpretation is complicated by the possible interference of the second representation based on the 21 momentum matrices. Also, the $v_{C,-D}$ part of the field must accompany the $(A,B)$ part and the role of the $(C,D)$ part in normalization issues and the effects, if any, on amplitudes need to be considered. Massive fields obtained with momentum matrices differ from those found without momentum matrices, bringing new terms for sources and motivating reinvestigation of interactions and gauge theory.

\appendix
\section{Spin and Momentum Matrices} \label{SMM}

The angular momentum and boost matrices of spin $(A,B)$ in a standard representation$^{ \cite{W6}}$ are
\begin{equation} \label{JAB+} (J^{\pm (A,B)})_{ab,a_1b_1} = (J^{(A,B)}_{x} \pm i J^{(A,B)}_{y})_{ab,a_1b_1} =  r^{(A)}_{\pm a_1} \delta_{a,a_1 \pm 1} \delta_{b, b_1} + r^{(B)}_{ \pm b_1} \delta_{a,a_1} \delta_{b, b_1 \pm 1}  \quad ,   \end{equation}
\begin{equation} \label{KAB+} (K^{\pm (A,B)})_{ab,a_1b_1} = (K^{(A,B)}_{x} \pm i K^{(A,B)}_{y})_{ab,a_1b_1} = -i( r^{(A)}_{\pm a_1} \delta_{a,a_1 \pm 1} \delta_{b, b_1} - r^{(B)}_{ \pm b_1} \delta_{a,a_1} \delta_{b, b_1 \pm 1} )  \quad ,  \end{equation}
\begin{equation} \label{JABz} (J_{z}^{(A,B)})_{ab,a_1b_1} = (a + b) \delta_{a,a_1} \delta_{b, b_1}    \quad , \end{equation}
\begin{equation} \label{KABz} (K_{z}^{(A,B)})_{ab,a_1b_1} = -i(a - b) \delta_{a,a_1} \delta_{b, b_1} \quad , \end{equation}
where
\begin{equation} \label{r}  r^{(A)}_{a_1} = \sqrt{(A - a_1)(A + a_1 + 1)} \quad . \end{equation}
Four types of spin $(A,B) \oplus (C,D)$  give nonzero momentum matrices:$^{ \cite{S}}$

\noindent Type 1: $A = C + 1/2; B = D + 1/2,$
\begin{equation} \label{pp+12} (P^{\pm}_{12})_{ab,cd} =  \frac{1}{2}(P_{x12} \pm iP_{y12})_{ab,cd} = \pm \frac{\sqrt{A \pm a}}{\sqrt{2A}}\frac{\sqrt{B \pm b}}{\sqrt{2B}} \enskip K^{12} \delta_{a,c \pm 1/2} \delta_{b,d \pm 1/2}  \quad , \end{equation}
\begin{equation} \label{ppz12}  \frac{1}{2}(P_{z12} \pm P_{t12})_{ab,cd} =  - \frac{\sqrt{A \pm a}}{\sqrt{2A}}\frac{\sqrt{B \mp b}}{\sqrt{2B}} \enskip K^{12}\delta_{a,c \pm 1/2} \delta_{b,d \mp 1/2} \quad , \end{equation}
\begin{equation} \label{pp+21} (P^{\pm}_{21})_{cd,ab} =  \frac{1}{2}(P_{x21} \pm iP_{y21})_{cd,ab} =  \pm \sqrt{A \mp a}\sqrt{B \mp b} \enskip K^{21} \delta_{c,a \pm 1/2} \delta_{d,b \pm 1/2}  \quad , \end{equation}
\begin{equation} \label{ppz21}  \frac{1}{2}(P_{z21} \pm P_{t21})_{cd,ab} =  + \sqrt{A \mp a}\sqrt{B \pm b} \enskip K^{21} \delta_{c,a \pm 1/2} \delta_{d,b \mp 1/2} \quad   \quad ,   \end{equation}

\noindent Type 2: $A = C + 1/2; B = D - 1/2,$
\begin{equation} \label{pm+12} (P^{\pm}_{12})_{ab,cd} =  \frac{1}{2}(P_{x12} \pm iP_{y12})_{ab,cd} =  \frac{\sqrt{A \pm a}}{\sqrt{2A}}\sqrt{D \mp d} \enskip K^{12} \delta_{a,c \pm 1/2} \delta_{b,d \pm 1/2}  \quad , \end{equation}
\begin{equation} \label{pmz12}  \frac{1}{2} (P_{z12} \pm P_{t12})_{ab,cd} =  \pm \frac{\sqrt{A \pm a}}{\sqrt{2A}}\sqrt{D \pm d} \enskip K^{12}\delta_{a,c \pm 1/2} \delta_{b,d \mp 1/2}   \quad ,  \end{equation}
\begin{equation} \label{pm+21} (P^{\pm}_{21})_{cd,ab} =  \frac{1}{2}(P_{x21} \pm iP_{y21})_{cd,ab} =   \sqrt{A \mp a}\frac{\sqrt{D \pm d}}{\sqrt{2D}} \enskip K^{21} \delta_{c,a \pm 1/2} \delta_{d,b \pm 1/2}  \quad , \end{equation}
\begin{equation} \label{pmz21}  \frac{1}{2}(P_{z21} \pm P_{t21})_{cd,ab} = \mp \sqrt{A \mp a} \frac{\sqrt{D \mp d}}{\sqrt{2D}} \enskip K^{21}\delta_{c,a \pm 1/2} \delta_{d,b \mp 1/2} \quad , \end{equation}

\noindent Type 3: $A = C - 1/2; B = D + 1/2,$
\begin{equation} \label{mp+12} (P^{\pm}_{12})_{ab,cd} =  \frac{1}{2}(P_{x12} \pm iP_{y12})_{ab,cd} =    \sqrt{C \mp c}\frac{\sqrt{B \pm b}}{\sqrt{2B}} \enskip K^{12} \delta_{a,c \pm 1/2} \delta_{b,d \pm 1/2}  \quad , \end{equation}
\begin{equation} \label{mpz12}   \frac{1}{2}(P_{z12} \pm P_{t12})_{ab,cd} =  \mp \sqrt{C \mp c}\frac{\sqrt{B \mp b}}{\sqrt{2B}} \enskip K^{12}\delta_{a,c \pm 1/2} \delta_{b,d \mp 1/2} \quad , \end{equation}
\begin{equation} \label{mp+21} (P^{\pm}_{21})_{cd,ab} =  \frac{1}{2}(P_{x21} \pm iP_{y21})_{cd,ab} =    \frac{\sqrt{C \pm c}}{\sqrt{2C}}\sqrt{B \mp b} \enskip K^{21} \delta_{c,a \pm 1/2} \delta_{d,b \pm 1/2}  \quad , \end{equation}
\begin{equation} \label{mpz21}  (P_{z21} \pm P_{t21})_{cd,ab} =  \pm \frac{\sqrt{C \pm c}}{\sqrt{2C}}\sqrt{B \pm b} \enskip K^{21}\delta_{c,a \pm 1/2} \delta_{d,b \mp 1/2}  \quad , \end{equation}

\noindent Type 4: $A = C - 1/2; B = D - 1/2,$
\begin{equation} \label{mm+12} (P^{\pm}_{12})_{ab,cd} =  \frac{1}{2}(P_{x12} \pm iP_{y12})_{ab,cd} =   \pm \sqrt{C \mp c}\sqrt{D \mp d} \enskip K^{12} \delta_{a,c \pm 1/2} \delta_{b,d \pm 1/2}  \quad , \end{equation}
\begin{equation} \label{mmz12}  \frac{1}{2}(P_{z12} \pm P_{t12})_{ab,cd} =   + \sqrt{C \mp c}\sqrt{D \pm d} \enskip K^{12}\delta_{a,c \pm 1/2} \delta_{b,d \mp 1/2}   \quad , \end{equation}
\begin{equation} \label{mm+21} (P^{\pm}_{21})_{cd,ab} =  \frac{1}{2}(P_{x21} \pm iP_{y21})_{cd,ab} =  \pm \frac{\sqrt{C \pm c}}{\sqrt{2C}}\frac{\sqrt{D \pm d}}{\sqrt{2D}} \enskip K^{21} \delta_{c,a \pm 1/2} \delta_{d,b \pm 1/2}  \quad , \end{equation}
\begin{equation} \label{mmz21}  \frac{1}{2}(P_{z21} \pm P_{t21})_{cd,ab} =  - \frac{\sqrt{C \pm c}}{\sqrt{2C}}\frac{\sqrt{D \mp d}}{\sqrt{2D}} \enskip K^{21}\delta_{c,a \pm 1/2} \delta_{d,b \mp 1/2}  \quad . \end{equation}

The angular momentum and boost matrices for spin  $(A,B) \oplus (C,D)$ are given in the text, (\ref{JABJCD}). Combining these angular momentum and boost matrices with the $P_{12}$ momentum matrices provides the 10 generators of a finite dimensional non-unitary matrix representation of the Poincar\'{e} algebra. Combining the angular momentum and boost matrices with the $P_{21}$ momentum matrices gives a second representation of the Poincar\'{e} algebra.

\section{Problems} \label{Pb}

\noindent 1. In order that spin $(A,B) \oplus (C,D)$ have nontrivial translation matrices, what spins $(C,D)$ can accompany (a) $(A,B)$ = $(0,1/2)$ (neutrino) and (b) $(A,B)$ = $(1/2,1/2)$ (photon)? (c) What is the helicity of the massless field for each choice?

\vspace{0.3cm}
\noindent 2. Show that the $u^{\prime}_{l}$ and $u_{l}$ in (\ref{up0}) and (\ref{up1}) satisfy (\ref{Du2}) .

\vspace{0.3cm}
\noindent 3. For displacement $\epsilon$ parallel to the standard momentum $k,$ i.e. $\epsilon^{\mu}$ = $\alpha k^{\mu},$ show that
$$   u_{l}(x,{\mathbf{p}},\sigma_{\pm}) = \sqrt{\frac{k^0}{p^0}} e^{i p \cdot x} D_{lm}(L(p) \bar{S} ,y) {e_{\pm}}_{m} \quad , $$
where $y^{\mu}$ = $x^{\mu} + \alpha p^{\mu}$ and $\bar{S}$ = $\bar{W}\bar{R}^{-1}$ is the translation matrix in the special little group transformation $\bar{W}.$ Also find a corresponding expression for $u^{\prime}_{l}(x,{\mathbf{p}},\sigma_{\pm},\Lambda,b)$

\vspace{0.3cm}
\noindent 4. Find the negative energy intertwiners and the transformed intertwiners $v_{l}(x,{\mathbf{p}},\sigma)$ and $v^{\prime}_{l}(x,{\mathbf{p}},\sigma,\Lambda,b)$ for helicities $\sigma_{\pm}$ = $A-B \pm 1$ and $\sigma_{0}$ = $A-B ,$ see reference \cite{W4} to set up the negative energy field construction.


\begin{thebibliography}{9}
\bibitem{W5} See, e.g., Weinberg, S., op. cit., p. 72. (Note that $L_1$ and $L_2$ are $A$ and $B$ in this reference.)
\end{thebibliography}
\end{document}